\documentclass[prl,twocolumn,superscriptaddress,showpacs]{revtex4}
\usepackage{graphicx}
\usepackage{amsmath}
\usepackage{amssymb}
\usepackage{color}

\renewcommand{\k}{\mathbf{k}}
\renewcommand{\r}{\mathbf{r}}

\begin{document}

\title{Impurity Bound States and Greens Function Zeroes as Local Signatures of Topology}
\author{Robert-Jan Slager}
\affiliation{Institute-Lorentz for Theoretical Physics, Leiden University, PO Box 9506, Leiden, The Netherlands}
\author{Louk Rademaker}
\affiliation{Kavli Institute for Theoretical Physics, University of California Santa Barbara, CA 93106, USA}
\author{Jan Zaanen}
\affiliation{Institute-Lorentz for Theoretical Physics, Leiden University, PO Box 9506, Leiden, The Netherlands}
\author{Leon Balents}
\affiliation{Kavli Institute for Theoretical Physics, University of California Santa Barbara, CA 93106, USA}
\date{\today}

\begin{abstract} 

We show that the local in-gap Greens function of a band insulator $\mathbf{G}_0 (\epsilon,\k_\parallel,\r_\perp=0)$, with $\r_\perp$ the position perpendicular to a codimension-1 or -2 impurity, reveals the topological nature of the phase. For a topological insulator, the eigenvalues of this Greens function attain zeros in the gap, whereas for a trivial insulator the eigenvalues remain nonzero. This topological classification is related to the existence of in-gap bound states along codimension-1 and -2 impurities. Whereas codimension-1 impurities can be viewed as 'soft edges', the result for codimension-2 impurities is nontrivial and allows for a direct experimental measurement of the topological nature of 2d insulators.
\end{abstract}

\pacs{71.10.Fd, 73.20.Hb }

\maketitle

The topological characterization of condensed states of matter has emerged as a prominent research interest over the last few decades. The flourishing of the Quantum Hall effect (QHE) \cite{Klitzing1980} in particular elucidated many connections between physical signatures and topological invariants \cite{tknn1982}, which supplement the order parameters of the usual symmetry-breaking Landau-Ginzburg paradigm. More recently, however, it became apparent that topological order can also arise by {\em virtue} of symmetry, and in particular the very common and robust Time Reversal (TR) symmetry is sufficient to establish the existence and stability of topological insulators\cite{KaneMele,Moore2007,Fu2007}.  This is quantified via a $\mathbb{Z}_{2}$ invariant, and results in gapless helical edge states or chiral Dirac fermions localized at the perimeter of the sample, in two and three dimensions, respectively.  The topological insulator has proven extremely rich both experimentally and theoretically\cite{hasan-kane-review,qi-zhang-review}.  The concept has been generalized to a ``periodic table'' describing various discrete symmetries and dimensions \cite{schnyder2008,schnyderNJP2010,kitaev2009}.   Lattice symmetries can similarly lead to further topological distinctions, result in ``crystalline topological insulators''\cite{Fu2011}, for which a general classification has been provided \cite{Nat-Phys-us2013}.

One may ask whether the topology of band insulators has some local signature? In fact, in this Letter, we will show that even the fully \emph{local} in-gap Greens function contains information about the band topology, which is then directly accessible by experiments.  The natural way this insight arises is through the study of impurities\cite{Lu:2011,Chang:2012,Black-Schaffer:2012}, similar to how the space group classification can be probed using lattice defects\cite{barkeshli-qiprx2012, barkeshli-qi-prb2013, andrejran2013, teokane2010,us2014, ranNatPhys2009, us-prl2012, NPB-us}. Consider a codimension-1 impurity line or surface in an insulator. In the limit where the impurity strength diverges, $V \to\infty$, such an impurity acts like a real edge which, following the bulk-boundary correspondence, should host zero gap metallic bound states in the topological phase. For finite $V$ the codimension-1 impurity surface can thus be viewed as a "soft edge".   The codimension-2 impurity lines or points do not host gapless states in the strong $V$ limit, so a priori there is no reason to expect they probe topology.  However, we will see that they in fact inherit the topological structure of the "soft edges". This nontrivial result implies that by probing bound states around a point impurity in a two-dimensional insulator, one can experimentally distinguish between the topological and trivial phase. 

Mathematically, the theory for impurity bound states shows that the existence of bound states in the topological phase is directly related to zero eigenvalues of the local in-gap Greens function. Consequently, we propose that the presence or absence of zero eigenvalues in the local in-gap Greens function is a signature of the band topology.

The remainder of this Letter is organized as follows. We first introduce the model and the theory of impurity bound states. Then we show that the local in-gap Greens function, relevant for codimension-1 and -2 impurities, has zero eigenvalues if and only if the phase is topological. We then propose an experiment that directly probes the Greens function zeroes for a point impurity in a 2d insulator. Finally, we relate our results to the known $\mathbb{Z}_{2}$ classification and the bulk-boundary correspondence.

{\it Theory of impurity bound states-}
We begin with a translationally invariant system, described by a minimal time-reversal invariant two-band model. The generic Hamiltonian assumes the form
\begin{equation}
	H_0 = \sum_{\k \alpha \beta} c^\dagger_{\k \alpha} d_\mu(\k) \gamma^\mu_{\alpha \beta} c_{\k \beta}
	\label{eq::Hamilton}
\end{equation}
where $\gamma^\mu$ are the $4 \times 4$ Dirac gamma matrices satisfying a Clifford algebra. We choose $\gamma^0 = \sigma_0 \otimes \tau_3$ and $\gamma^i = \sigma_i \otimes \tau_1$. Here, the $\sigma$ and $\tau$ Pauli matrices act in the spin and orbital space, respectively.  Time reversal (TR) symmetry then implies that $d_0(\k)$ must be even and $d_i(\k)$ must be an odd function.
In particular, we focus on the representative cases that $d_0(\k) = M - 2B \sum_{i} (1 - \cos k_i)$ and $d_i(\k) = \sin k_i$ \cite{bhz, 3da, 3db}. This is the familiar class of models displaying topological nontrivial regimes for parameter range $0 < M/B < 4d$. This specific choice is not expected to restrict our results, as a topological insulator generically has by adiabatic continuity the form of a lattice regularized massive Dirac Hamiltonian. Moreover, extra terms that respect TR symmetry will not change the results described below.  

The real frequency Greens function in the gap reads
\begin{equation}
	\mathbf{G}_0 (\omega, \k) = \frac{1}{\omega - d_\mu(\k) \gamma^\mu}
	= \frac{\omega \mathbf{1}_4 + d_\mu(\k) \gamma^\mu}{\omega^2 - |d(\k)|^2}.
	\label{GreensF}
\end{equation}
Subsequently, we introduce an impurity into the system, which in general can be described by the Hamiltonian
\begin{equation}
	H_V = \sum_{\r \alpha \beta} c^\dagger_{\r \alpha} V_{\alpha \beta}(\r) c_{\r \beta}.
\end{equation}
To find the corresponding spectrum in the presence of the impurity, one needs to solve the (differential) Schr\"{o}dinger equation
\begin{equation}
	(H_0 + H_V) \psi_\epsilon (\r) = \epsilon \psi_\epsilon (\r)
	\label{SE}
\end{equation}
where $\epsilon$ is the energy of the state. This can be transformed into an integral equation, \cite{Mahan}
\begin{equation}
	\psi_\epsilon (\r) = \sum_{\r'} \mathbf{G}_0 (\epsilon, \r - \r') V(\r') \psi_\epsilon (\r').
	\label{IntegralEquation}
\end{equation}
For an insulator, the real-frequency Greens function is well-behaved inside the gap and displays exponential decay as a function of $\r$. This implies that an in-gap solution of Eqn. (\ref{IntegralEquation}) will yield a bound state around the impurity.

The existence of an in-gap bound state depends on $H_0$ and the shape of the impurity potential $V(\r)$. However, the qualitative difference between topological and trivial band insulators is found in Greens function features, and is therefore largely independent of the choice of impurity potential. Let us therefore consider the simplest possible choice: a constant $V(\r)$ along a $n$-dimensional plane in a $d$-dimensional system (hence codimension $d-n$). The $d$-dimensional position vector $\r$ can be split into the perpendicular coordinates $\r_\perp$ and the parallel coordinates $\r_\parallel$, so that the impurity potential is given by
\begin{equation}
	V(\r) = \mathbf{V}_0 \delta^{n}_{\r_\perp = 0} 
\end{equation}
where we have used a Kronecker delta to signify our use of lattice models and introduced the $4\times4$ Hermitian matrix $\mathbf{V}_{0}$. We note that even if the potential $V(r)$ just couples directly to the electron density, $\mathbf{V}_{0}$ is not necessarily diagonal when expressed in terms of the second quantized operators. Explicitly, $\psi(\r)$ is the electron field and can be expanded in orbital wave functions as $\psi(\r) = \sum_{i \alpha} c_{i \alpha} \phi_{i \alpha}(\r)$, where $\phi_{i \alpha}(\r)$ is the wave function of the $\alpha$-orbital at lattice site $i$ satisfying the orthonormality condition $\int d^d\r \phi^*_{i \alpha} (\r) \phi_{j \beta} (\r) = \delta_{ij} \delta_{\alpha \beta}$.  Hence, the total density is diagonal, $N = \int d^d \r \psi^\dagger(\r) \psi(\r) = \sum_{i\alpha} c^\dagger_{i \alpha} c_{i \alpha}$. However, an impurity that couples to the density, generally does not give rise to a diagonal expression,
\begin{eqnarray}
	H_V &=& \int d^d \r V(\r) \psi^\dagger(\r) \psi(\r)\\ \nonumber
		&=& \sum_{i \alpha \beta} c^\dagger_{i \alpha} c_{i \beta} \int d^d\r V(\r) \phi^*_{i \alpha} (\r) \phi_{i \beta} (\r)\nonumber
\end{eqnarray}
The shape of $\mathbf{V}_0$ can be restricted, though, using symmetry principles. For example, when we consider nonmagnetic impurities, TR invariance applies to the impurity potential as well \footnote{In fact, one can show that magnetic impurities will result in a doubling of the impurity bound state solutions, see the Appendix.}. As a result, the matrix $\mathbf{V}_{0}$ has only six degrees of freedom,
\begin{equation}
	\mathbf{V_0} =  V \mathbf{1} + V_0 \sigma^0 \otimes \tau^3
		+ V_i \sigma^i \otimes \tau^2 + V_4 \sigma^0 \otimes \tau^1.
\end{equation}
Recall that in this notation, time reversal is $T = i \sigma^2 K$ where $K$ is complex conjugation. Parity, on the other hand, is given by $\gamma^0 = \sigma^0 \otimes \tau^3$. If we require both parity and TR  the form of $\mathbf{V}_{0}$ is even further constrained to
\begin{equation}
	\mathbf{V}_0 =  V \mathbf{1} + V_0 \gamma^0.
\end{equation}
Since the translational symmetry is not broken along directions parallel to the impurity, the impurity bound states have a well-defined parallel momenta $\k_\parallel$. Thus $\psi_\epsilon(\r) \propto e^{i \k_\parallel \cdot \r_\parallel}$ and the integral equation Eqn. (\ref{IntegralEquation}) reduces to an eigenvalue equation for each $\k_\parallel$,
\begin{equation}
	\det \left[ \mathbf{G}_0 (\epsilon,\k_\parallel,\r_\perp=0) \mathbf{V}_0 - \mathbf{1} \right] = 0.
	\label{EvEquation}
\end{equation}
We immediately notice that for the case $\mathbf{V}_0 = V\mathbf{1}$, the existence of bound states is directly related to the eigenvalues of the local ($\r_\perp = 0$) in-gap Greens function $\mathbf{G}_0 (\epsilon,\k_\parallel,\r_\perp=0)$.

{\it Codimension-1 impurities-}
Let us now consider codimension-1 impurities, that is a surface in $d=3$ and a line in $d=2$, having only one perpendicular direction $r_\perp = 0$, see Eqn. (\ref{EvEquation}). The corresponding Greens function in the gap, integrated over the perpendicular momentum, can be decomposed in terms of 
\begin{eqnarray}
	g_\mu (\epsilon, \k_\parallel ) & = & \int \frac{dk_\perp}{2\pi}
		\frac{d_\mu(\k_\parallel,k_\perp)}{\epsilon^2 - |d(\k_\parallel,k_\perp)|^2}, \\
	g (\epsilon, \k_\parallel) & = & \int \frac{dk_\perp}{2\pi}
		\frac{\epsilon}{\epsilon^2 - |d(\k_\parallel,k_\perp)|^2}, 
\end{eqnarray}
so that $\mathbf{G_0} (\epsilon,\k_\parallel,r_\perp=0) = g (\epsilon,\k_\parallel) \mathbf{1} + g_\mu (\epsilon, \k_\parallel ) \gamma^\mu$. 

At any TR-symmetric point for the parallel momentum, for example $\k_\parallel = 0$ or $\pi$, the $g_\parallel$ are vanishing. Additionally, $g_\perp$ vanishes since the integrand is an odd function of $\k_\perp$. At TR-symmetric points we thus only need to consider $g_0$ and $g$. We note that this still holds if we add e.g. Rashba spin orbit coupling terms, that are odd functions of the momentum, to the bare Hamiltonian. This allows to us to verify the following results also in the absence of any other symmetry but TR symmetry. 

\begin{figure}
	\includegraphics[width=0.8\columnwidth]{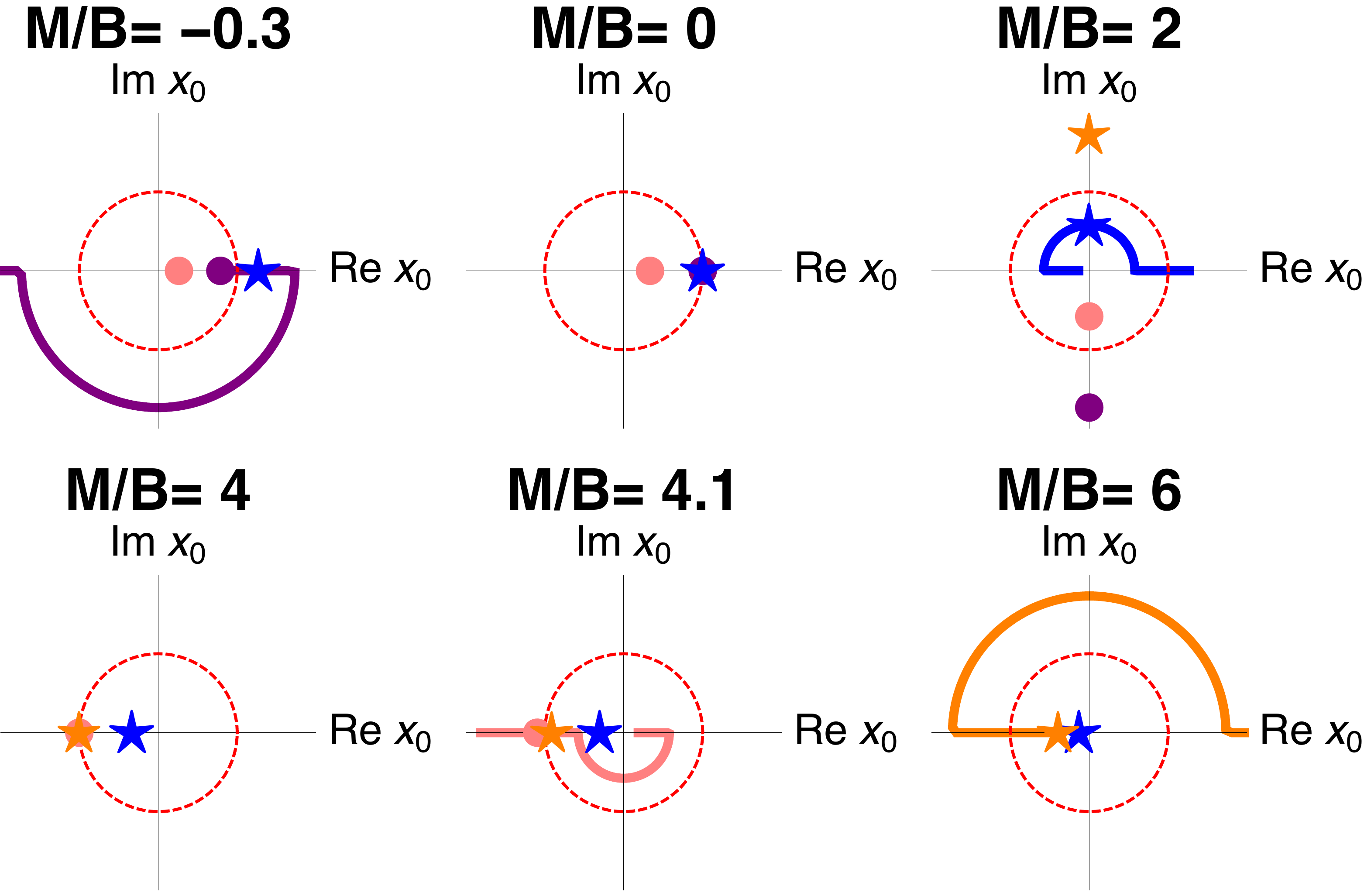}
	\caption{The flow of the poles of Eq. \eqref{Poles} as function of $M/B$. The two negative residue poles are indicated with a dot, whereas the stars mark the poles having a positive residue. In the trivial system, two poles with either positive or negative residue are located inside the red unit circle. For $M/B=0,4$ there are two poles with opposite residue located {\em on} the unit circle, signaling the transition. The topological phase is characterized by two poles of opposite residue in the unit circle. }
	\label{FigPoles}
\end{figure}

\begin{figure}
	\includegraphics[width=\columnwidth]{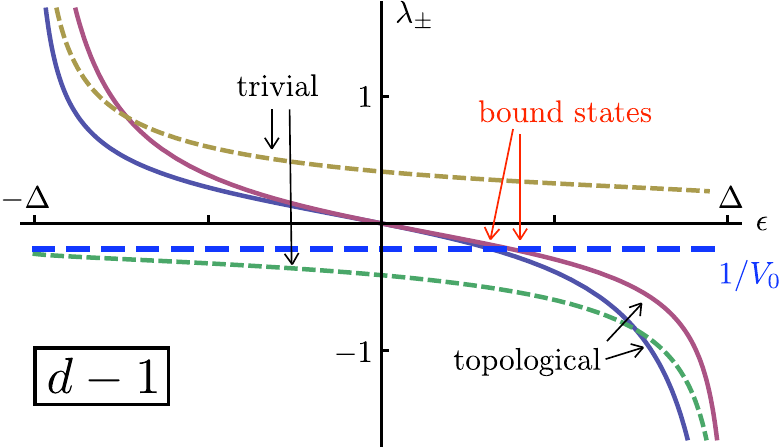}
	\caption{The eigenvalues of the Greens function \mbox{$\mathbf{G}_0(\epsilon,k_\parallel=0, r_\perp=0)$} in the two and three dimensional $M/B$ model, relevant for codimension-1 defects. In the trivial system (dashed lines, $M/B=-1$) the eigenvalues $\lambda_\pm$ are nonzero, which means for strong enough impurities there are no in-gap bound state possible. In the topological system (solid line, $M/B=1$) the eigenvalues cross zero energy at $\epsilon=0$ and hence in-gap bound states always exist.}
	\label{1din2d2}
\end{figure}

We will now show that the eigenvalues of the local in-gap Greens function $\mathbf{G_0} (\epsilon,\k_\parallel^S,r_\perp=0)$,
\begin{equation}
	\lambda_\pm (\epsilon) = g(\epsilon) \pm g_0 (\epsilon),
\end{equation}
have the shape displayed in Fig. \ref{1din2d2}. The difference between a topological insulator and a trivial insulator, is whether the in-gap Greens function has a zero eigenvalue or not. Consequently, this result implies that for each choice of impurity potential $\mathbf{V_0}$, a topological insulator will have at least one in-gap bound state.

Note that the denominator in the integrand, $\epsilon^2 - |d|^2$, is always negative. In the trivial phase $d_0 (\mathbf{k})$ does not change sign throughout the Brillouin zone. This implies that $g_0(\epsilon)$ does not change sign. Additionally, because $\epsilon < | d_0 (\k)|$, we have $g(\epsilon)+ |g_0(\epsilon)| > 0$ and $g(\epsilon) - |g_0(\epsilon)| <0$. Therefore, in the trivial phase, the Greens function $\mathbf{G_0} (\epsilon,\k_\parallel, \r_\perp = 0) $ \emph{never} has an eigenvalue equal to zero \emph{for all} momenta $\k_\parallel$ and energies $\epsilon$. 

On the other hand, in the topological phase, when $\k_\parallel$ is chosen as a TR-symmetric point $\k_{S}$ associated with the projection describing the topological phase \cite{Nat-Phys-us2013}, the Greens function satisfies $\mathbf{G_0} (\epsilon=0, \k_\parallel = \k_S, \r_\perp=0) = 0$. To prove this, we evaluate
 \begin{eqnarray}
	g_0(0,\k_{S}) &=& -\int_{-\pi}^\pi \frac{dk}{2\pi} \frac{d_0(k, \k_{S})}{|d^2(k, \k_{S})|} 
		\nonumber \\
		&=& -\int_{-\pi}^\pi \frac{dk}{2\pi}\nonumber
		\frac{\hat{M}(\k_{S})-2B(1-\cos k)}{\sin^2 k + (\hat{M}(\k_{S})-2B(1-\cos k))^2},
	\label{EQ1d}
\end{eqnarray}
where $\hat{M}(\k_{S})=M-2B\sum_{i=1}^{d-1}(1-\cos(k_i))$ in terms of the coordinates $\k_{S}$.
Substituting $x = e^{ik}$, the integral becomes a contour integral over the unit circle and we obtain
\begin{equation}
	= \frac{1}{2\pi i}
	\oint_{|x|=1}dx
	\frac{ B + (\hat{M}-2B)x + Bx^2 }
	{f(x)}
\end{equation}
with $f(x) = \left(\frac{1}{2} + 6B^2 - 4B\hat{M} + \hat{M}^2 \right) x^2
	+ 2B (\hat{M}-2B) x(x^2+1)
	+ \left( B^2 - \frac{1}{4} \right) (x^4+1)$. 
This expression is then solved by application of the Cauchy residue theorem. The poles are located at, see Fig. \ref{FigPoles},
\begin{equation}\label{Poles}
	x_{0} = \frac{ 2B - \hat{M} \pm \sqrt{1-4B\hat{M}+\hat{M}^2}}{2B \pm 1}.
\end{equation}
When the system becomes gapless for $\hat{M}=0$ or $\hat{M}=4B$, we find that there are two zero's located {\em on} the unit circle. Thus, from the analytic structure of the integral we infer that the regime $0<\hat{M}<4B$ is topologically distinct from $\hat{M}<0$ and $4B<\hat{M}$. Focusing on the topological phase $0<\hat{M}<4B$, we observe that the residues of the two zeroes  $x_{0,\pm}= \frac{ 2B - \hat{M} \pm \sqrt{1-4B\hat{M}+\hat{M}^2}}{2B + 1}$ located inside the unit circle
cancel,
\begin{equation}
	\mathrm{Res}(x_{0,+})
		+\mathrm{Res}(x_{0,-})
	= 0.
\end{equation}
Therefore, in the topological phase, $g_0(\epsilon=0) = 0$. When $\hat{M}<0$ or $4B<\hat{M}$, that is in the trivial phase, the integral never equates to zero. Together with the universal divergence of the both $g(\epsilon)$ and $g_0(\epsilon)$ at the band edges as shown in the Appendix, we arrive at the generic description as shown in Fig. \ref{1din2d2}. In the topological phase the Greens function switches sign and thus has at least one energy $\epsilon$ for which it has a zero eigenvalue.

Consequently, for any impurity strength a topological insulator will always have in-gap states along a codimension-1 impurity, whereas for a trivial insulator it depends on specific details of the impurity and the insulator. The codimension-1 impurity can thus be understood as a 'soft edge'.

{\it Codimension-2 impurities-}
The above results on the structure of the eigenvalues of $\mathbf{G_0}(\epsilon,\k_\parallel,\r_\perp=0)$ in the codimension-1 case can be extended to codimension-2 impurities. In this case, there are two perpendicular directions $\k_\perp = (k^x_\perp ,k^y_\perp)$,
\begin{eqnarray}
	\mathcal{G}_\mu (\epsilon, \k_\parallel) & = & \int \frac{dk^x_\perp dk^y_\perp}{(2\pi)^2}
		\frac{d_\mu(\k_\parallel,\k_\perp)}{\epsilon^2 - |d(\k_\parallel,\k_\perp)|^2}, \\
	\mathcal{G} (\epsilon, \k_\parallel) & = & \int \frac{dk^x_\perp dk^y_\perp}{(2\pi)^2}
		\frac{\epsilon}{\epsilon^2 - |d(\k_\parallel,\k_\perp)|^2}, 
\end{eqnarray}
for $\mu=0,1,2,3$. It is clear that for any of the perpendicular directions $\mathcal{G}_\perp = 0$, as the integrand is odd. 

For the trivial phase, we can show that the eigenvalues of $\mathbf{G_0}(\epsilon,\k_\parallel,\r_\perp=0)$ are nonzero throughout the gap, since the two-dimensional integral can be done by first integrating in one direction, which yields the results from the codimension-2 impurities, and then integrating along the second direction. Therefore, $\mathbf{G_0}(\epsilon,\k_\parallel,\r_\perp=0)$ is never zero in the gap.

\begin{figure}
	\includegraphics[width=\columnwidth]{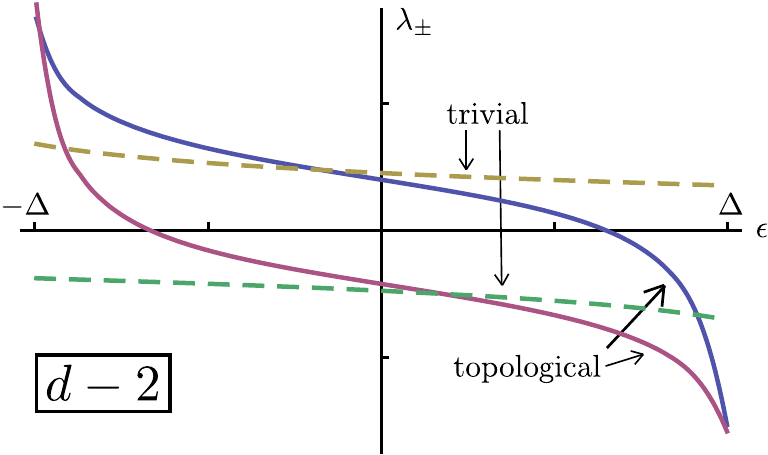}
	\caption{The in-gap eigenvalues of the local Greens function \mbox{$\mathbf{G}_0(\epsilon, r=0)$} in the two dimensional $M/B$ model. In the trivial system (dashed lines, $M/B=-1$) the eigenvalues are nonzero, whereas for the topological system (solid line, $M/B=1$) each eigenvalue is zero for some energy.}
	\label{d-2Imp}
\end{figure}

Of more interest is the question of existence of zero energy eigenvalues in the topological regime. Let us focus on the  $2$-dimensional case, so that there are no parallel directions: we are directly probing the \emph{local, on-site Greens function}. We expect that the terms $\mathcal{G}(\epsilon)$ and $\mathcal{G}_0(\epsilon)$ will diverge close to the band-edge. In fact, these divergences are captured by expanding around the point where the gap is minimal, $\k_G$,
\begin{equation}
	|d(k^x,k^y)|^2 = \Delta^2 + a (k^x-k^x_G)^2 + b(k^y-k^y_G)^2 + \ldots
\end{equation}
The diverging part of the integral is then captured by the integral
\begin{eqnarray} \label{div1}
	\int - \frac{dk^x dk^y}{(2\pi)^2} \frac{1}{ 2 \Delta \delta \epsilon + a (k^x)^2 + b (k^y)^2 }\\
	\sim - \int_0^{0^+} \frac{dq}{ 2\pi \sqrt{ab}} \frac{q}{2 \Delta \delta \epsilon + q^2}
	\sim \frac{ \log \delta \epsilon}{4 \pi \sqrt{ab}}.
\end{eqnarray}
Hence, $\mathcal{G}(\epsilon) \sim  \frac{- |\Delta| \log \delta \epsilon}{4 \pi \sqrt{ab}}$ and $\mathcal{G}_0(\epsilon) \sim  \frac{d_0(\k_G)  \log \delta \epsilon}{4 \pi \sqrt{ab}}$ in proximity of the valence band. The dependence of the gap $\Delta$ on $d_0(\k)$ proves that both eigenvalues in the topological phase diverge to positive infinity at the valence band edge, and to minus infinity at the conduction band edge. Consequently, in the topological phase the Greens function eigenvalues must be zero somewhere in the gap. Details are provided in the Appendix. This proves that in $d=2$, the completely local in-gap Greens function $\mathbf{G_0}(\epsilon,r=0)$ has zero eigenvalues if and only if the system is in the topological phase, see Fig. \ref{d-2Imp}.

This result carries over to the case of line impurities in $d=3$ topological insulators, if the remaining parallel momentum is chosen at one of the TR-symmetric points.

\begin{figure}
	\includegraphics[width=\columnwidth]{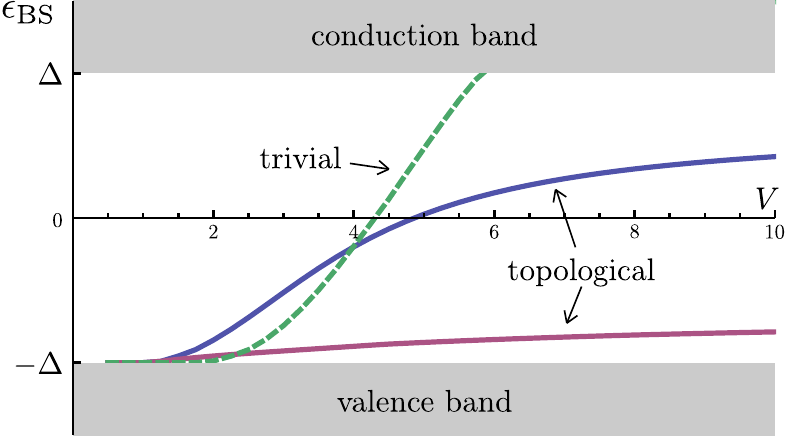}
	\caption{Typical energy of impurity bound states in a two-dimensional insulator as a function of impurity strength $V$. Here, we discern between the topological regime (solid lines, $M/B=1$) and the trivial regime (dashed line, $M/B=-1$). For strong $V$, bound state in the trivial phase has disappeared in the conduction band, whereas the bound states in the topological phase remain.}
	\label{EBS}
\end{figure}

{\it Experiment-} The existence of these zero eigenvalues can be probed directly in experiments, using impurity bound states as solutions to Eqn. (\ref{EvEquation}). Imagine a two-dimensional insulator, where at one isolated point a tunable gate voltage is applied, serving as the impurity potential $V$. Then using tunneling spectroscopy, the possible bound states around this impurity can be found. Upon \emph{increasing} the impurity potential $V$, the energies of the bound states shift: for a trivial insulator, one can make a bound state disappear into one of the bands by a sufficiently strong potential. However, our results show that for a topological insulator, for all strong $V$ there will always be two bound states. Explicitly, the energy of the bound state as a function of $V$ is shown in Fig. \ref{EBS}.

{\it Classification and the bulk-boundary correspondence-}
The odd number of crossings per spin branch of $\mathbf{G_0} (\epsilon,\k_\parallel, \r_\perp = 0)$ with the zero eigenvalue axis is a topological property \cite{ZubkovVolovik,Zubkov, Gurari}. It reflects that in the nontrivial regime the system has an odd number of Kramers degenerate edge states on either side of the surface and hence may be regarded as a consequence of the bulk-boundary correspondence \cite{hasan-kane-review, qi-zhang-review}. 
In particular, the bulk TR $\mathbb{Z}_{2}$ invariant is in this case simply the product $\prod_{\Gamma_{i}}\xi_{i}$ of the parity $\gamma_{0}$ eigenalues $\xi_{i}$ over the TR points in the Brillouin zone \cite{fuKane}. Moreover, the two relevant poles have the same residues but multiplied by sign of the mass, i.e. the parity $\gamma_{0}$ eigenvalues. Hence, only if the choice $\k_{S}$ in the projected plane is associated with two masses of opposite sign, meaning that this cut features a band inversion, the poles cancel in the above integral rendering a zero eigenvalue.
This is in accordance with the space group classification \cite{Nat-Phys-us2013}. 

For example, for $d=2$ the model Eqn. (\ref{eq::Hamilton}) exhibits a $\Gamma$ ($T-p4mm$) phase for $0<M/B<4$ and a $M$ ($T-p4$) phase for $4<M/B<8$. From the above considerations we find that $\hat{M}$ has to satisfy $0<\hat{M}<4$ for  $\mathbf{G_0} (\epsilon,\k_\parallel, \r_\perp = 0)$ to develop zero eigenvalues. Taking subsequently projections onto $k_{x}$ and $k_{y}$ and using that $\hat{M}(\k_{s})=M-2B\sum_{i}(1-\cos(k_i))$, we thus conclude that in the $\Gamma$ phase these $k_{S}$ choices correspond to an inversion at $\k =(0,0)$, whereas in the $M$ phase the inversion is at $\k =(\pi,\pi)$. We stress that this analysis still holds if we add Rashba terms, that are odd functions of momentum. Similarly, we may add a next nearest neighbor term $\tilde{d}_0(\k)=-\tilde{B}[1-\cos(k_{x})\cos(k_{y})], d_x(\k) = \cos(k_{x})\sin(k_{y}), d_y(\k) = -\sin(k_{x})\cos(k_{y})$ to the Hamiltonian \cite{Nat-Phys-us2013}, allowing for an additional $X-Y$ ($p4$) topological crystalline phase \cite{Fu2011}. This phase is associated with the inversion momenta $\k =(\pi,0)$ {\em and} $\k =(0,\pi)$. An identical calculation then shows that indeed projections alongs $k_{x}$ and $k_{y}$ yield zero eigenvalues of $\mathbf{G_0} (\epsilon,\k_\parallel, \r_\perp = 0)$ for {\em both} $k_S = 0$ and $\pi$. These ideas carry over directly to three dimensions. Consider for example a $\Gamma$ ($T-pm\bar{3}m$) phase with an inversion at $\k=(0,0,0)$ for $0<M/B<4$, in the projected plane one should now choose $\k_S = (0,0)$ or on a line $k_S = 0$. 

{\it Conclusions and Outlook-}
We have shown that topological band insulators can be characterized by the existence of zero eigenvalues in the local in-gap Greens function $\mathbf{G}_0(\epsilon,\k_\parallel,\r_\perp = 0)$, where $\r_\perp$ is the position vector perpendicular to a codimension-1 or -2 impurity. Whereas the codimension-1 impurities can be viewed as soft edges, the nontrivial result for codimension-2 impurities suggests one can experimentally probe the difference between a topological insulator and a trivial insulator using a tunable localized impurity.

We made some simplifying assumptions in the proof presented above, but the results are robust.  For example, adding more bands to the system, further away from the Fermi level, might introduce extra impurity bound states whose energies depend strongly on the impurity strength, but one can show that these do not generally remove the states arising from the low energy bands.   Furthermore, we showed (e.g. Fig.~\ref{EBS}) the persistence of bound states for strong potential $V$ but neglected $V_0$;  however, while the shape of bound state energy versus potential strength changes if $V_0$ is included, the conclusion that they persist is independent of the ratio $V_0/V$.   Furthermore, the principle of adiabatic continuation suggests our proposed classification applies equally well to gapped interacting systems with a quasiparticle description\cite{Raghu2008} or topological superconductors\cite{qi-zhang-review}.

{ \it Acknowledgments-}
This work is supported by the Dutch Foundation for Fundamental Research on Matter (FOM). L.R. was supported by the Dutch Science Foundation (NWO) through a Rubicon grant.  L.B. was supported by the National Science Foundation under grant No.DMR-12-06809.

\newpage

\newpage

\appendix

\section{Appendix}
\subsection{Details codimension-1 case}
We have shown in the main text that  $\mathbf{G_0} (\epsilon=0, \k_\parallel = \k_S, \r_\perp=0) = 0$ only acquires zero eigenvalues in the topological regime. Here, we elaborate on the generic structure which results in the schematic representation as conveyed in Fig. \ref{1din2d2}. Here we address the behavior of $\mathbf{G_0}$ close to the band edges. First, note that $g(\epsilon)$ is a decreasing odd function in $\epsilon$, and $g_0(\epsilon)$ is an even function in $\epsilon$. Close to the band edges, both $g(\epsilon)$ and $g_0(\epsilon)$ will always diverge as a square root, $g \sim 1/\sqrt{\delta \epsilon}$. Whenever the gap is located at a $T$-symmetric point, where $d_i(\k) = 0$, the strengths of the divergences in $g(\epsilon)$ and $g_0(\epsilon)$ are equal. Without loss of generality, to prove this we choose $m<\frac{1}{2}$ where the gap is located at $\k = 0$. We expand $|d(\k_\parallel=0,k_\perp)|^2$ around that point,
\begin{equation}
	|d(\k_\parallel=0,k_\perp)|^2 = m^2 + (1 - 2m) k_\perp^2 + \ldots.
\end{equation}
The divergence of $g(\epsilon)$ near the valence band edge can be isolated by integrating only over a small region $(-\alpha,\alpha)$ around the top of the valence band for small $\delta \epsilon = \epsilon + |m|$,
\begin{eqnarray}
	g(\epsilon = -|m| + \delta \epsilon) &\approx&
		\int_{-\alpha}^\alpha \frac{dk_\perp}{2\pi}
		\frac{|m|}{2 |m| \delta \epsilon + (1 - 2m) \k_\perp^2} \\
	&\approx&
		\sqrt{\frac{|m|}{8 (1-2m) \delta \epsilon}} + \mathcal{O}(\delta \epsilon^0).
		\label{Divergence}
\end{eqnarray}
A similar argument applies to $g_0(\epsilon)$ close to the valence band, as the magnitude of the gap $|d_0(\k_\parallel = 0, k_\perp = 0)| = |m|$. Therefore $g_0(\epsilon)$ and $g(\epsilon)$ have exactly the same divergent behavior close to the band edge. Hence, $g + |g_0|$ only diverges close to the valence band and $g-|g_0|$ diverges close to the conduction band. Note that this analysis is valid for $m<\frac{1}{2}$, when the band gap is at $\k=(0,0)$, and for $m>\frac{15}{2}$, when the band gap is at $\k=(\pi,\pi)$. This implies that this cancellation of divergences is present in the trivial phase (which supports the notion that there the eigenvalues do not change sign), but also for some region in the topological phase.
When $m>1/2$ and $m<15/2$, the gap is not located at a $T$-symmetric point but rather at $(0,k_G)$ or $(\pi,k_G)$. In this case, the gap satisfies $\Delta^2 = d_0(k_G)^2 + \sin^2 k_G > d_0(k_G)^2$. This last one is important, because then at any $T$-symmetric point of $\k_\parallel$, the eigenvalues of $\mathbf{G_0}(\epsilon,\k_\parallel^S)$ will both diverge to positive infinity at the valence band edge. To see that, we expand
\begin{equation}
	|d(\k)|^2 = \Delta^2 + a (k_\perp - k_G)^2 + \ldots
\end{equation}
so that the divergent parts close to the valence band edge of the Greens function terms are
\begin{eqnarray}
	g(\epsilon = -|\Delta| + \delta \epsilon) & \sim & \frac{|\Delta|}{\sqrt{8a |\Delta| \delta \epsilon}} \\
	g_0(\epsilon = -|\Delta| + \delta \epsilon) & \sim & \frac{-d_0(0,k_G)}{\sqrt{8a |\Delta| \delta \epsilon}} .
\end{eqnarray}
Because $|\Delta| > |d_0|$, we find that divergences do not cancel and both eigenvalues of $\mathbf{G_0}(\epsilon,\k_\parallel)$ will diverge to positive infinity at the valence band edge. This means that at the conduction band edge, both diverge to negative infinity, therefore implying an energy at which $\mathbf{G_0}(\epsilon)$ has zero eigenvalues.

Finally, note that the eigenvalues of $\mathbf{G_0} (\epsilon,\k_\parallel^S,r_\perp=0) \mathbf{V_0}$, relevant for the impurity problem, are given by
\begin{equation}
	\lambda_{GV}= V_0 g_0 + V g \pm
		\sqrt{ (Vg_0 + V_0g)^2 + \left(\sum_{i=1}^4 V_i^2 \right) (g^2 - g_0^2)}
	\label{ImpEqnd1}
\end{equation}
and are thus directly dependent on the eigenvalues of $\mathbf{G_0} (\epsilon,\k_\parallel^S,r_\perp=0)$.

\subsection{Magnetic codimension-1 impurities}
Let us shorty address the fate of codimension-1 magnetic impurities. In this case, TR invariance does not constrain the specific structure of the $V$-matrix. Let us, however, focus on the simplest case: $\mathbf{V_0} \sim \sigma^i \otimes \tau^0$. We assume the impurity line orientation to be the $2$-direction, so that in $d=2$ we have 
\begin{equation}
	\mathbf{G_0} (\epsilon,\k_\parallel,r_\perp=0) = g(\omega) \mathbf{1} + g_0 (\omega) \gamma^0 + g_2 (\omega) \gamma^2.
\end{equation}
In addition, we assume the following impurity shape
\begin{equation}
	\mathbf{V_0} =  \sum_{i=1}^2 V_i \sigma^i \otimes \tau^0.
\end{equation}
which differentiates between the parallel $V_2$ and perpendicular $V_1$ magnetization directions. The resulting structure has some nice analytic properties. In particular, it follows from 
\begin{widetext}
$\lambda_{GV} (\epsilon) = \pm \sqrt{(g^2+g_0^2)(V_1^2+V_2^2) + g_2^2 (V_2^2 - V_1^2)\pm 2 \sqrt{ (g^2 (V_1^2 +V_2^2) - g_2^2 V_1^2) (g_0^2 (V_1^2 +V_2^2) + g_2^2 V_2^2)}}$,
\end{widetext}
that for every eigenvalue $\lambda_{GV}(\epsilon)$ there is also an eigenvalue $- \lambda_{GV} (\epsilon)$. Therefore, if $\lambda_{GV}(\epsilon)=1$ for a given $\epsilon$, then there is also a solution at $-\epsilon$. We thus conclude that impurity bands resulting from magnetic impurities are symmetric in energy. This is corroborated with explicit tight-binding results in final section of this Appendix.

In the special case where the magnetic impurity is aligned to the impurity, effectively setting $V_{1}$ to zero, the eigenvalues simplify as
\begin{equation}
	\lambda_{GV} (\epsilon) = ( \pm g \pm \sqrt{g_0^2 + g_2^2}) V_2
\end{equation}
which is just $\pm1$ times the solution for the identity matrix impurity, $\mathbf{V_0} \sim \mathbf{1}$. Hence, in these cases one just retrieves a doubling of the results found in the case of nonmagnetic impurities.

\subsection{Details codimension-2 impurities}
We here corroborate the ideas of the main text with the specific arguments, showing that $\mathbf{G_0}(\epsilon,\k_\parallel,\r_\perp=0)$ in two dimensions has zero eigenvalues in the topological regime.
From Eq. \eqref{div1} , it follows that $\mathcal{G}(\epsilon) \sim  \frac{- |\Delta| \log \delta \epsilon}{4 \pi \sqrt{ab}}$ and $\mathcal{G}_0(\epsilon) \sim  \frac{d_0(\k_G)  \log \delta \epsilon}{4 \pi \sqrt{ab}}$ close to the valence band. If the gap is not at a $T$-symmetric point, this automatically implies that the divergences do not cancel and we are left with eigenvalues that all diverge at both band edges, leading to the fact that the Greens functions eigenvalues have to be zero somewhere in the gap.

On the other hand, if the gap is at a symmetric point we need an extra argument to show the result of Fig. \ref{d-2Imp}, hence consider $0<m<\frac{1}{2}$. Trivially, $\mathcal{G}(\epsilon) > 0$ for $\epsilon<0$ and $\mathcal{G}(\epsilon=0) = 0$. At the same time, we know that both $\mathcal{G}_0(\epsilon)$ and $\mathcal{G} (\epsilon)$ have a logarithmic divergence at the valence band.  However, the difference is that because $d_0(k=0)>0$, the function $\mathcal{G}_0(\epsilon)$ goes to minus infinity. At the same time, $\mathcal{G}(\epsilon)$ goes to plus infinity because in that case the numerator is $\epsilon < 0$. Thus the two lines must cross if $\mathcal{G}_0(\epsilon=0)>0$, and consequently there is a point where $\mathbf{G_0}$ has zero eigenvalues.

It remains to prove that, in this construction, $\mathcal{G}_0(\epsilon=0) > 0$ for $0<m<\frac{1}{2}$. This actually straightforward. Since $\mathcal{G}_0(\epsilon=0) = \int \frac{dk_\parallel}{2\pi} g_0 (\epsilon=0,k_\parallel)$ we directly infer that
\begin{equation}
	g_0 (\epsilon=0,k_\parallel) = (m - 4 + 2 \cos k_\parallel) \int \frac{dk_\perp}{2\pi} \frac{-1}{|d(k_\parallel,k_\perp)|^2} > 0
\end{equation}
for all $k_\parallel$ given $m<2$, and hence $\mathcal{G}_0(\epsilon=0) > 0$ in the desired region $0<m<\frac{1}{2}$.

\subsection{Tight-binding results}
\begin{figure*}
	\includegraphics[width=0.9\textwidth]{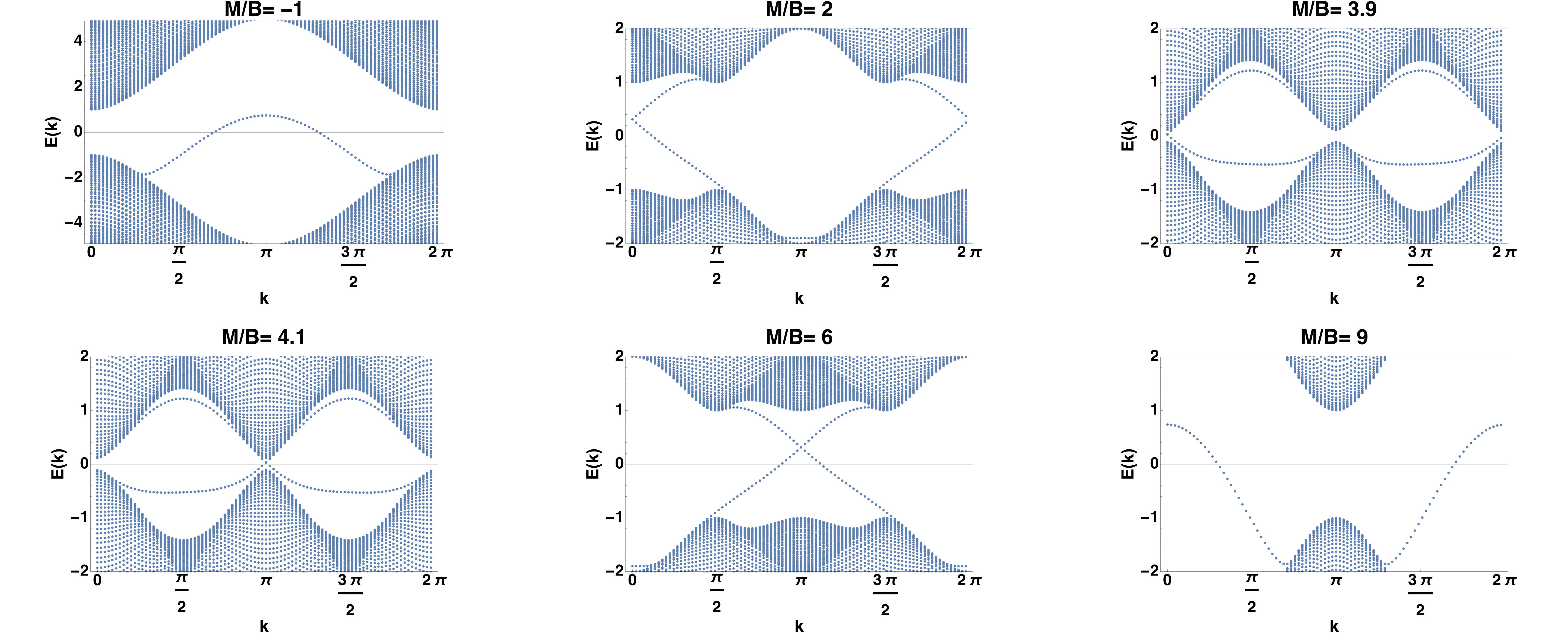}
	\caption{In gap impurity bands for various values of $M/B$ for a impurity line that hosts a delta potential having strength $\mathbf{V}_{0}/B=-6\times\mathbf{1}$. Specifically, these simulations were performed on a two dimensional $60\times90$ system with periodic boundary conditions. }
	\label{V=-6Plot}
\end{figure*}

\begin{figure*}
	\includegraphics[width=0.9\textwidth]{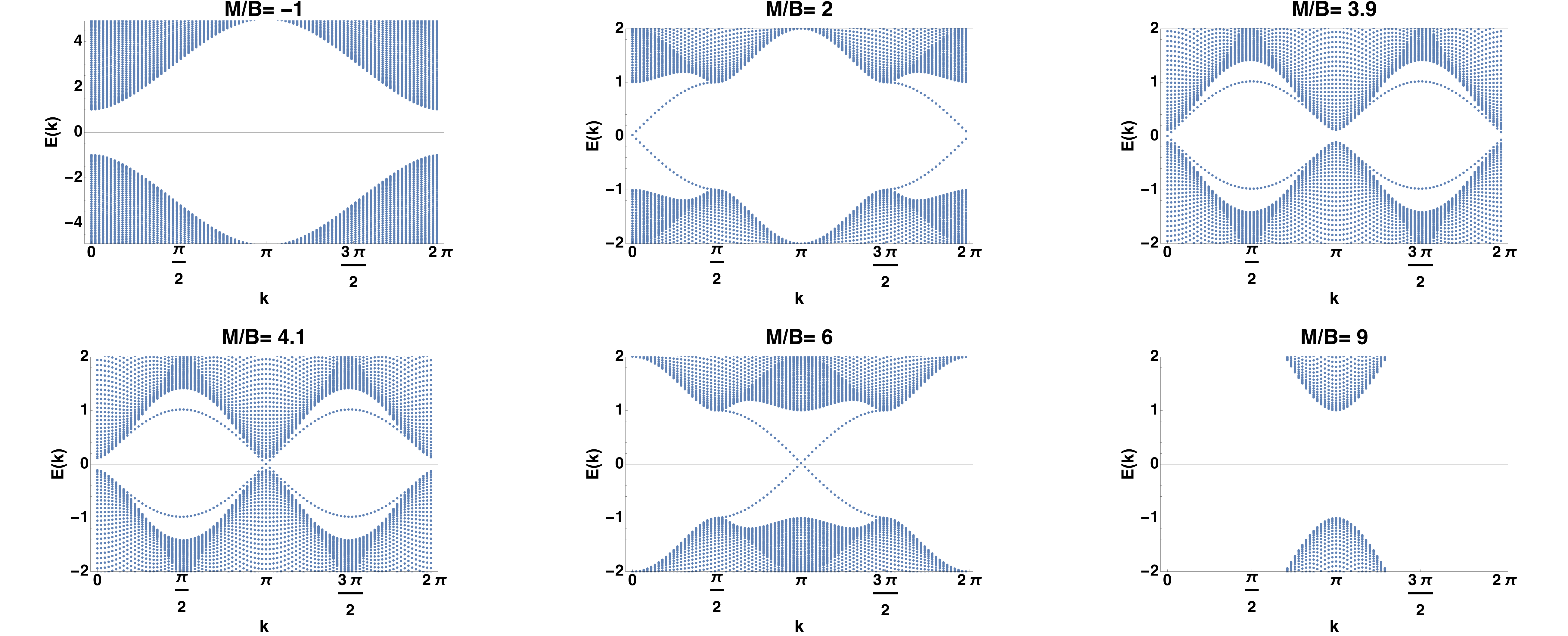}
	\caption{Analogous plot as the previous one, but now for $\mathbf{V}_{0}/B=-100\times\mathbf{1}$. We observe that in the gap there are only impurity bands for the topological regime. In addition, a difference between the $\Gamma$ phase ($0<M/B<4$) and $M$ phase ($4<M/B<8$) can be discerned. Namely, in the first case the bands are related to the inversion of the bulk bands at $\k=0$ and the bands are symmetric around the project $k=0$ value. Similarly, the $M$ phase is associated with $\k=(\pi,\pi)$.}
	\label{V=-100Plot}
\end{figure*}

The main consequence of the difference in structure of $\mathbf{G}_0 (\epsilon,\k_\parallel,\r_\perp=0)$ in the trivial and topological regime is reflected in the formation of bound states in the gap between the two relevant topologically active bands. As this provides for a derived experimental signature, let us therefore corroborate the results with explicit tight-binding calculations. Fig. \ref{V=-6Plot} shows the results for the two dimensional $M-B$ model  in presence of a codimension-1 impurity given by the identity matrix with strength $-6$ in units of $B$. The formed impurity bands in the gap of the valence and conduction band are clearly visible. When the impurity strengths are subsequently increased, to $V=-100$, only the topological regime still hosts impurity bands as demonstrated in Fig. \ref{V=-100Plot}. This is in accordance with the results in the main text. 

\begin{figure*}
	\includegraphics[width=0.9\textwidth]{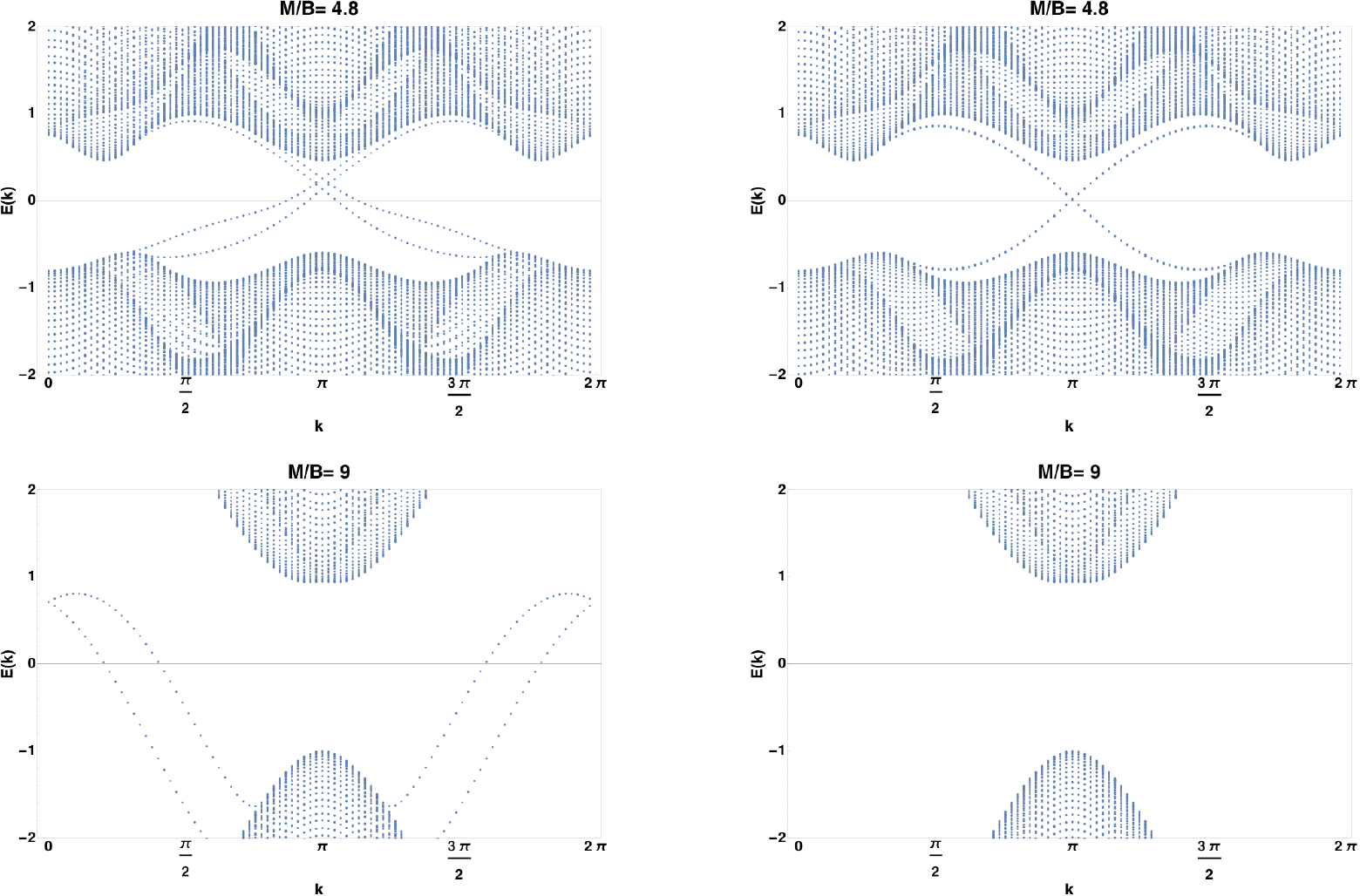}
	\caption{Spectra in the presence of a Rashba term, that break the particle hole symmetry of the original Hamiltonian. The strength of the Rashba coupling equates in all cases to $R_{0}/B=0.6$. The resultant effect is the lifting of the degeneracies of the impurity bands. Nonetheless, the qualitative features relate to the outlined results and previous figures. Specifically, the panels on the left show the system in presence of an impurity potential 
		$\mathbf{V}_{0}/B=-6\times\mathbf{1}$, whereas the right have $\mathbf{V}_{0}/B=-100\times\mathbf{1}$. }
	\label{RashbaPlot}
\end{figure*}

The inclusion of Rashba type couplings to the Hamiltonian that do not close the band gap,
\begin{equation}
	H_{\mathrm{Rashba}} = \sum_{\k} R_{o} 
	\frac{1}{2} (1+\tau_{z})
	(\sigma_{y} \sin k_{x}-\sigma_{x} \sin k_{y}),\end{equation}
results in the lifting of the degeneracy of the impurity band as shown. Nonetheless, the qualitative results are the same. Some representative cases connecting with the previous results are summarized in Fig. \ref{RashbaPlot}.

\begin{figure*}
	\includegraphics[width=0.9\textwidth]{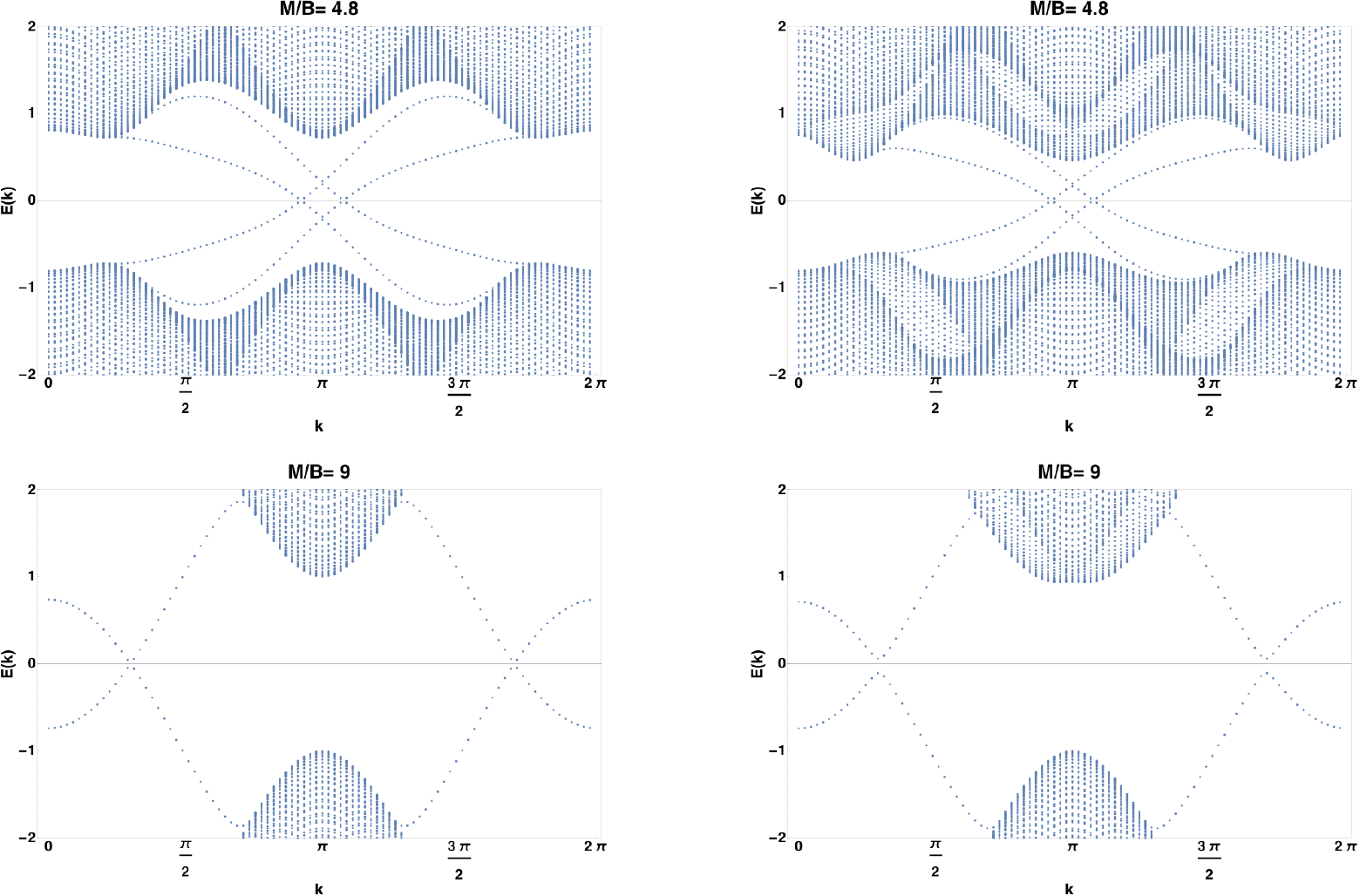}
	\caption{Impurity bands originating from the inclusion of a codimension-1 magnetic impurity.  In all cases the impurity strength $V/B$ equated to $-6$. The left panels show the magnetic result in absence of Rashba terms.The right panels display the same information in presence of a Rashba terms with coupling strength $R_{0}/B=0.6$. We observe that the bands are symmetric around zero energy. Interestingly, the degeneracy of the nodes is lifted in the trivial regime but not in topological regime. The instances shown are for the case of a parallel magnetic impurity, i.e $\mathbf{V_0} =  V_i \sigma^i \otimes \tau^0$ with $i$ the direction of the line impurity. The results for perpendicular realizations resulted in the exact spectra.}
	\label{MagneticPlot}
\end{figure*}

Finally, the last Fig. \ref{MagneticPlot} shows the spectrum in the presence of magnetic impurities. As indicated by the above analysis one finds in this case symmetric impurity bands. Also in this case Rashba terms do not affect the outlined analysis.

\end{document}